\documentstyle[aps,multicol]{revtex}

\begin{document}
\draft

\title{
Parity effect in a small superconducting grain:
A rigorous result
}
\author{Guang-Shan Tian$^{(1,2)}$
and Lei-Han Tang$^{(1)}$}
\address{
$^{(1)}$ Department of Physics, Hong Kong Baptist University, 
Kowloon Tong, Kowloon, Hong Kong\\
$^{(2)}$ Department of Physics, Peking University, Beijing 100871, China}

\date{\today}
\maketitle
\begin{abstract}
The parity effect in an ultra-small superconducting grain is examined.
By applying a generalized
version of Lieb's spin-reflection positivity technique,
we show rigorously that the parity parameter $\Delta_P$
is nonvanishing in such a system.
A positive lower bound for $\Delta_P$ is derived.
\end{abstract}

\begin{multicols}{2}

\section{Introduction}

Recent experiments by Ralph, Black and Tinkham
\cite{Ralph,Black}
on tunneling currents through nanometer-scale Al grains have rekindled
theoretical interest in modifications of the 
original Bardeen-Cooper-Schrieffer (BCS) theory\cite{Bardeen} 
when the average level spacing $\delta\epsilon$ 
becomes comparable to the bulk superconducting gap $\Delta$. 
Although the possible failure of the BCS mean-field theory
in this regime was speculated by Anderson\cite{Anderson} in 1959,
quantitative studies of the crossover from ultra-small grain
to bulk behavior emerged only recently
\cite{Janko,Delft,Smith,Matveev,Braun1,Braun2,Mastellone,Berger,Balian}.

A natural generalization of the superconducting gap
$\Delta$ for a bulk superconductor to the case of
a small grain with a fixed
number of electrons is the parity parameter
\begin{equation}
\Delta_P \equiv E_0(2N + 1) - \frac{1}{2}
[E_0(2N) + E_0(2N + 2)]
\label{Parameter}
\end{equation}
where $E_0(M)$ is the ground state energy of a grain of $M$ electrons.
Jank\'o, Smith, and Ambegaokar\cite{Janko} first discussed the parity
dependence of the superconducting gap in a small superconductor and
showed that $\Delta_P$ reduces to the bulk superconducting
gap $\Delta$ when the level spacing $\delta\epsilon\rightarrow 0$.
Using a path integral technique, Matveev and Larkin\cite{Matveev}
calculated $\Delta_P$ for the BCS Hamiltonian 
[see Eq. (\ref{Hamiltonian1})], and found a nonvanishing result also 
in the ultra-small grain limit $\delta\epsilon\gg \Delta$.
This is in contrast with previous mean-field calculations
\cite{Delft,Smith} which yielded a vanishing superconducting
gap when the grain becomes sufficiently small.
Recent numerical investigations\cite{Mastellone,Berger} 
on specific models in the region
$\delta\epsilon\approx\Delta$ have confirmed the conclusions 
of Matveev and Larkin, although a full analytical solution
of the problem even for the simplest type of models in this class is
not yet available.

The purpose of this paper is to present some rigorous results
on the positivity of the parity parameter $\Delta_P$ under
somewhat more general conditions than those considered
so far in the literature. By applying the recently developed
Lieb's spin-reflection positivity technique \cite{Lieb1,Lieb2,Lieb3}, 
we show that $\Delta_P$ is strictly positive in the BCS model
of superconductivity for any value of
$\delta\epsilon/\Delta$ and an arbitrary distribution of the single-particle
energy levels $\epsilon_k$ (see below).
In addition, we derive a lower bound for $\Delta_P$ which is independent
of the number of particles in the grain.

The paper is organized as follows. In Sec. II we introduce the
model and collect some of the known symmetry properties of the
Hamiltonian considered. In Sec. III we introduce a generalized
version of Lieb's spin-reflection positivity method and present
a proof for the strict positivity of the parity parameter $\Delta_P$.
Specializing on the commonly studied case of a constant pair
coupling constant, we derive a positive lower bound for $\Delta_P$.
Sec. IV contains a summary of our main results and discussions
on the further possible extensions.

\section{The Hamiltonian and its basic symmetries}

As a plausible description of electronic states in an isolated
grain of a superconductor ${\cal M}$, let us consider
the BCS pairing Hamiltonian\cite{Bardeen}
\begin{equation}
H_{\cal M} =
\sum_{k,\>\sigma} \epsilon_k
c_{k\sigma}^\dagger c_{k\sigma} -
\sum_{\mid\epsilon_k\mid,\>\mid\epsilon_{k^\prime}\mid < \omega_D}
g(k,\>k^\prime) c_{k\uparrow}^\dagger c_{k\downarrow}^\dagger
c_{k^\prime\downarrow} c_{k^\prime\uparrow}.
\label{Hamiltonian1}
\end{equation}
Here $k$ labels the single-particle eigenstates
$\mid k\rangle$ of $\cal M$ with energies $\epsilon_k$, and
$c_{k\sigma}^\dagger$ ($c_{k\sigma}$)
is the fermion creation (annihilation) operator
which creates (annihilates) a fermion of spin
$\sigma$ in state $\mid k\rangle$.
The cut-off energy for the pairing interaction is given by
$\omega_D>0$, which, for conventional superconductors,
can be taken to be the Debye frequency. The sums in (\ref{Hamiltonian1})
extend over all {\it admissible} states $\mid k\rangle$ of ${\cal M}$
which satisfy the condition $\mid\epsilon_k\mid<\omega_D$.
The total number of such states is denoted by $N_E$.
The coupling constants $g(k,\>k^\prime)$
are assumed to be {\it positive} for all pairs of admissible states.
In the Hamiltonian (\ref{Hamiltonian1}),
only those matrix elements of the interaction responsible for 
superconductivity are included. The contributions of the other
terms are assumed to be negligible. In particular, no 
pair breaking terms are present.

The average energy level spacing $\delta\epsilon=1/{\cal N}(\epsilon_F)$ 
in a metallic grain of diameter $d$ is proportional to $1/d$, 
where ${\cal N}(\epsilon_F)$ is the density of states at the Fermi level.
In the bulk limit $d\rightarrow\infty$, the BCS mean-field theory
applies and a superconducting ground state, characterized
by a finite superconducting gap 
\begin{equation}
\Delta\simeq 2\omega_D\exp(-\delta\epsilon/g),
\label{BCS-gap}
\end{equation}
is obtained.
However, when the grain size is reduced, $\delta\epsilon$ increases
and eventually becomes comparable or larger than the bulk superconducting gap
$\Delta$, at which point quantum fluctuations in the 
occupation numbers need to be taken into account.

The analysis presented in this paper is for a grain with a
fixed number of particles. It is therefore important to note
the following symmetries of the Hamiltonian (\ref{Hamiltonian1}),

(i) $H_{\cal M}$ commutes with the total fermion number operator
$\hat{N}=\sum_k\left(\hat n_{k\uparrow}+\hat n_{k\downarrow}\right)$, 
where 
$\hat n_{k\sigma}=c_{k\sigma}^\dagger c_{k\sigma}$
is the fermion number operator on state $\mid k\rangle$ with spin $\sigma$.
Consequently,
the total Hilbert space of the Hamiltonian
decomposes into numerous invariant subspaces
$\{V(M)\}$. Each of them is characterized
by the particle number $M$.

(ii) Next, consider the total spin operators of the
grain $\cal M$,
\begin{eqnarray}
\hat{S}_+ &\equiv& \sum_{k }c_{k\uparrow}^\dagger
c_{k\downarrow},\>\>\>
\hat{S}_- \equiv \sum_{k} c_{k\downarrow}^\dagger
c_{k\uparrow},\nonumber\\
\hat{S}_z &\equiv& \frac{1}{2}
\sum_{k} \left(\hat n_{k\uparrow} -
\hat n_{k\downarrow}\right).
\label{Spin Operators}
\end{eqnarray}
Then, $H_{\cal M}$ also commutes with these operators.
Therefore, both the total spin $S$ and the $z$-component
$S_z$ are good quantum numbers.
In addition, for a given $M$ and $S$, states with different $S_z$
are degenerate.

(iii) Another important property of the Hamiltonian (\ref{Hamiltonian1}) 
is that it does not contain pair-breaking or single-particle
hopping terms. Thus, in constructing the eigenstates of the
Hamiltonian, one can further restrict oneself to a subspace
with a fixed proportion of paired and unpaired particles.
In this subspace, the unpaired particles go into a fixed
set of single-particle eigenstates, while the paired particles
may choose to be in any of the remaining states.
This property greatly simplifies the numerical procedure
\cite{Mastellone,Berger}
to find the ground state of the Hamiltonian (\ref{Hamiltonian1}),
although in the analysis presented in this paper we do not need
to presume such behavior.

In Refs.~\cite{Matveev}, \cite{Mastellone}
and \cite{Berger}, the case of uniform level spacing
$\epsilon_k=k\delta$ was considered.
Furthermore, $g(k,\>k^\prime)=g>0$ is
assumed to be a constant for any pair of admissible states
$\mid k\rangle$ and $\mid k^\prime\rangle$.
In the following, these conditions will be relaxed.
We shall only require that $\{\epsilon_k\}$ are {\it real
quantities} and $g(k,\>k^\prime)>0$ for any pair
of admissible states $\mid k\rangle$ and $\mid k^\prime\rangle$.

\section{The parity parameter}

\subsection{Matrix representation of wavefunction}

Our result on the positivity of the parity parameter $\Delta_P$
is derived from a special representation of eigenstate
wavefunctions previously used in the study of the binding energy of 
fermions in the negative-$U$ Hubbard model\cite{Tian1}
with an odd number of particles.
The idea originates from Lieb's spin-reflection positivity
method\cite{Lieb1,Lieb2,Lieb3} for a system with equal number of up and
down spins. In the following we repeat some
of the essential steps leading to the representation and refer the
reader to the literature for additional details.

By regrouping the fermion operators, we can rewrite the
Hamiltonian (\ref{Hamiltonian1}) as
\begin{eqnarray}
H_{\cal M} &=& \sum_{k,\>\sigma} \epsilon_k
c_{k\sigma}^\dagger c_{k\sigma}
\nonumber\\
&&- \sum_{\mid\epsilon_k\mid,\>\mid\epsilon_{k^\prime}\mid<\omega_D}
g(k,\>k^\prime) \left(c_{k\uparrow}^\dagger
c_{k^\prime\uparrow}\right)
\left(c_{k\downarrow}^\dagger
c_{k^\prime\downarrow} \right)
\label{Hamiltonian2}
\end{eqnarray}
Notice that the Hamiltonian is symmetric with
respect to the up-spin and the down-spin fermion
operators (spin-reflection). However, this form is not 
particularly convenient
when we calculate matrix elements of $H_{\cal M}$
in an occupation number representation.
To deal with this problem, we shall consider the so-called
``pseudo-fermion'' operators, which were first introduced 
by Jank\'o {\it et al.} \cite{Janko} to
study systematically BCS superconductivity with fixed number parity
and then, re-introduced
by Lieb and Nachtergaele \cite{Lieb3} to show that the ground
state of the Hubbard model at half-filling is the global
ground state when the chemical potential $\mu=\frac{U}{2}$. 
These pseudo-fermion operators are defined by
\begin{equation}
\hat{C}_{k\uparrow} \equiv
c_{k\uparrow},\>\>\>
\hat{C}_{k\downarrow} \equiv
(-1)^{\hat{N}_\uparrow} c_{k\downarrow}
\end{equation}
where $\hat{N}_\uparrow$ is the number operator
of the up-spin fermions in the system.
Apparently, these operators satisfy the
conventional fermionic anticommutation relations
\begin{eqnarray}
\{\hat{C}_{k\sigma}^\dagger,\>
\hat{C}_{k^\prime\sigma}\}
&=& \delta_{kk^\prime},\nonumber\\
\\
\{\hat{C}_{k\sigma},\>
\hat{C}_{k^\prime\sigma}\} &=&
\{\hat{C}_{k\sigma}^\dagger,\>
\hat{C}_{k^\prime\sigma}^\dagger\} = 0.
\nonumber
\end{eqnarray}
However, the operators $\hat{C}_{k\uparrow}$ and
$\hat{C}_{k^\prime\downarrow}$ now {\it commute} with each other.
Consequently, Hamiltonian (\ref{Hamiltonian2})
can be rewritten in the following direct-product form,
\begin{eqnarray}
H_{\cal M}
= &&
\sum_{k,\>\sigma} \epsilon_k
\left(\hat{n}_{k\uparrow} \otimes \hat{I}_\downarrow +
\hat{I}_\uparrow \otimes \hat{n}_{k\downarrow} \right) \nonumber\\
&& - 
\sum_{\mid\epsilon_k\mid,\>\mid\epsilon_{k^\prime}|<\omega_D}
\left(\sqrt{g(k,k^\prime)}
\hat{C}_{k\uparrow}^\dagger
\hat{C}_{k^\prime\uparrow}\right) \nonumber\\
&&\qquad\qquad\otimes
\left(\sqrt{g(k,k^\prime)}
\hat{C}_{k\downarrow}^\dagger
\hat{C}_{k^\prime\downarrow}\right)
\label{Hamiltonian3}
\end{eqnarray}

Using the pseudo-fermion operators, we can also write
the wavefunction of a state 
with $N_\uparrow$ spin-up particles and $N_\downarrow$ spin-down
particles in the form,
\begin{equation}
\Psi = \sum_{m=1}^{D_\uparrow}
\sum_{n=1}^{D_\downarrow} W_{mn} \chi_m^\uparrow
\otimes \chi_n^\downarrow.
\label{Wave Function}
\end{equation}
In Eq.~(\ref{Wave Function}), $\chi_s^\sigma$ is a state vector
defined by
\begin{equation}
\chi_s^\sigma \equiv \hat{C}_{k_1\sigma}^\dagger
\cdots \hat{C}_{k_L\sigma}^\dagger \mid 0\rangle
\end{equation}
where $\left(k_1,\cdots,k_L\right)$,
$L=N_\uparrow$, for $\sigma=\uparrow$;
$L=N_\downarrow$, for $\sigma=\downarrow$,
denote the admissible states occupied
by the fermions with spin $\sigma$.
Apparently, the entire set $\{\chi_s^\sigma\}$
gives a natural basis for $V_\sigma(L)$,
the subspace of $L$ fermions with spin $\sigma$,
whose dimension is denoted by $D_\sigma(L)$.

For the case $D_\uparrow=D_\downarrow$, 
it can be shown (see below)
that the coefficient matrix ${\cal W}$ in (\ref{Wave Function})
can be brought into a diagonal form through (separate) unitary
transformations in the $V_\uparrow(N_\uparrow)$ and 
$V_\downarrow(N_\downarrow)$ subspaces, respectively. 
To generalize this property to the case 
$D_\uparrow\neq D_\downarrow$ or equivalently $N_\uparrow\neq N_\downarrow$
(which is the case for an odd number
of electrons in the grain), we consider a larger subspace
${\cal H}_\uparrow=V_\uparrow(N_\uparrow) \oplus V_\uparrow(N_\downarrow)$
and
${\cal H}_\downarrow=V_\downarrow(N_\uparrow)\oplus 
V_\downarrow(N_\downarrow)$,
for the spin-up and spin-down particles, respectively\cite{Tian1}.
This way the number of basis $D=D_\uparrow+D_\downarrow$ for 
spin-up and spin-down states becomes the same.
The original wavefunction (\ref{Wave Function}) can now be written 
in the form,
\begin{equation}
\Psi = \sum_{m,\>n} \tilde W_{mn} \chi_m^\uparrow
\otimes \chi_n^\downarrow,
\label{expand_wave}
\end{equation}
where 
\begin{equation}
\tilde{\cal W} = \left(
\begin{array}{lll}
O\>& {\cal W}\\
O\>& O
\end{array}
\right)
\label{Matrix}
\end{equation}
is a $D\times D$ {\it square matrix}.

Let us now introduce the polar factorization lemma in matrix theory.

{\bf Lemma:} Let $A$ be an {\it arbitrary} (not necessarily
Hermitian) $n\times n$ matrix.
Then, there are two $n\times n$ {\it unitary}
matrices $U$, $V$ and an $n\times n$
{\it diagonal} semi-positive definite matrix
$R$ such that
\begin{eqnarray}
& &
A=URV, \nonumber\\
& &
r_{mm'}=r_m\delta_{mm'}\>\>{\rm and}
\>\>r_m\ge 0,\>\>m=1,\cdots,n.
\end{eqnarray}

The proof of this lemma can be found in any standard
textbook of matrix theory\cite{Book} and is also presented in the appendix
of Ref.~\cite{Tian1}.

Applying this lemma to $\tilde{\cal W}=URV$, we can rewrite
(\ref{expand_wave}) as
\begin{eqnarray}
\Psi& = &
\sum_{m=1}^D \sum_{n=1}^D
\tilde W_{mn} \chi_m^\uparrow \otimes
\chi_n^\downarrow \nonumber\\
& = &
\sum_{m=1}^D \sum_{n=1}^D \left(URV\right)_{mn}
\chi_m^\uparrow \otimes \chi_n^\downarrow\nonumber\\
& = &
\sum_{l=1}^D
r_l \psi_l^\uparrow \otimes
\phi_l^\downarrow,
\label{WF2}
\end{eqnarray}
where $\{r_l\}$ is a set of nonnegative numbers and
\begin{equation}
\psi_l^\uparrow = \sum_{m=1}^D
U_{ml} \chi_m^\uparrow,\>\>\>\>
\phi_l^\downarrow = \sum_{n=1}^D
V_{ln} \chi_n^\downarrow.
\end{equation}

Since $U$ and $V$ are unitary,
$\{\psi_l^\uparrow\}$ and 
$\{\phi_l^\downarrow\}$ are also
orthonormal bases in subspaces
${\cal H}_\uparrow$
and ${\cal H}_\downarrow$, respectively. Furthermore,
since $\Psi$ is an eigenvector of
$\hat{N}_\uparrow$ and $\hat{N}_\downarrow$,
the following conditions should hold
\begin{equation}
\langle\Psi\mid \hat{N}_\uparrow
\mid\Psi\rangle = 
\sum_{l=1}^D r_{l}^2 \langle\psi_l\mid \hat{N}
\mid\psi_l\rangle = N_\uparrow,
\label{Constraint1}
\end{equation}
and
\begin{equation}
\langle\Psi\mid \hat{N}_\downarrow
\mid\Psi\rangle = 
\sum_{l=1}^D 
r_l^2\langle\phi_l\mid \hat{N}
\mid\phi_l\rangle = N_\downarrow,
\label{Constraint2}
\end{equation}
where we have used the normalization condition,
\begin{equation}
\langle\Psi\mid\Psi\rangle =\sum_{l=1}^D r_l^2=1.
\label{normalization}
\end{equation}
In both Eqs.~(\ref{Constraint1}) and (\ref{Constraint2}),
the spin indices are dropped in the sums, because,
in each equation, only one species of spin is involved.

It is useful to note that, although the transformations
$U$ and $V$ may mix states with different number of 
particles, those states $\psi_l^\uparrow$ and $\phi_l^\downarrow$ 
which actually appear in the sum (\ref{WF2}), i.e., 
those correspond to $r_l>0$, are {\it eigenstates} of the
number operators $\hat N_\uparrow$ and $\hat N_\downarrow$,
respectively. Thus,
\begin{equation}
\hat{N}\mid\psi_l\rangle= N_\uparrow\mid\psi_l\rangle,
\quad
\hat{N}\mid\phi_l\rangle= N_\downarrow\mid\phi_l\rangle,
\quad{\rm for}\>\> r_l>0.
\label{number_op}
\end{equation}
The proof is left as an exercise for the reader.

\subsection{Parity parameter in the ground state}

With these preparations, we now summarize our main result
in the following theorem.

{\bf Theorem:} Let $E_0(M)$ be the ground
state energy of Hamiltonian (\ref{Hamiltonian1}) with $M$
particles and $\Delta_P$ be the parity
effect parameter defined in Eq.~(\ref{Parameter}).
Then, $\Delta_P$ is {\it strictly} positive for any integer
$N$ subject to $N<N_E$.

{\it Proof:} To prove this theorem, 
let us now consider the ground state of a system
with $2N+1$ particles and a total spin $S$. 
Since $\hat{S}_+$ and $\hat{S}_-$
commute with $H_{\cal M}$, we can specialize on the case
$S_z=\frac{1}{2}\left(N_\uparrow-N_\downarrow\right)=\frac{1}{2}$,
i.e., $N_\uparrow=N+1$ and $N_\downarrow=N$.
Let $\mid\Psi_0(2N+1)\rangle$ 
be the ground state wavefunction
with the above properties. Using the representation (\ref{WF2})
for $\mid\Psi_0(2N+1)\rangle$ and the Hamiltonian (\ref{Hamiltonian3}),
we can write the ground state energy of $2N+1$ particles in the form,
\begin{eqnarray}
& &
E_0(2N + 1) \nonumber\\
& \equiv &
\langle\Psi_0(2N + 1)\mid H_{\cal M} \mid
\Psi_0(2N + 1)\rangle \nonumber\\
& = &
\sum_{l=1}^D r_l^2
\left[\langle\psi_l^\uparrow\mid
\hat{T}_\uparrow
\mid\psi_l^\uparrow\rangle +
\langle\phi_l^\downarrow\mid
\hat{T}_\downarrow
\mid\phi_l^\downarrow\rangle\right] \nonumber\\
& - &
\sum_{k,k^\prime,l_1,l_2}
r_{l_1}r_{l_2} \langle\psi_{l_2}^\uparrow\mid
\hat{Q}_\uparrow(k,k^\prime)\mid\psi_{l_1}^\uparrow\rangle
\langle\phi_{l_2}^\downarrow\mid
\hat{Q}_\downarrow(k,k^\prime)
\mid\phi_{l_1}^\downarrow\rangle,
\nonumber\\
\label{Energy}
\end{eqnarray}
where
\begin{equation}
\hat{T}_\sigma=\sum_k\epsilon_k \hat n_{k\sigma},\>\>\>
\hat{Q}_\sigma(k,\>k^\prime) = \sqrt{g(k,\>k^\prime)}
\hat{C}_{k\sigma}^\dagger \hat{C}_{k^\prime\sigma}
\end{equation}
By applying inequality
$\mid ab\mid\le\frac{1}{2}(\mid a\mid^2+\mid b\mid^2)$
to each term in the second sum of Eq.~(\ref{Energy}) and
dropping the spin indices, we obtain
\begin{eqnarray}
& &
E_0(2N + 1) \nonumber\\
& \ge &
\frac{1}{2} \sum_{l=1}^D r_l^2 \left[\langle\psi_l\mid
\hat{T} \mid\psi_l\rangle + \langle\psi_l\mid \hat{T}
\mid\psi_l\rangle \right] \nonumber\\
& + &
\frac{1}{2} \sum_{l=1}^D r_l^2
\left[\langle\phi_l\mid \hat{T} \mid\phi_l\rangle +
\langle\phi_l\mid \hat{T} \mid\phi_l\rangle\right] \nonumber\\
& - &
\frac{1}{2} \sum_{k,k^\prime,l_1,l_2}
r_{l_1} r_{l_2} \langle\psi_{l_2}\mid
\hat{Q}(k,k^\prime) \mid\psi_{l_1}\rangle
\overline{\langle\psi_{l_2}\mid
\hat{Q}(k,k^\prime) \mid\psi_{l_1}\rangle}
\nonumber\\
& - &
\frac{1}{2} \sum_{k,k^\prime,l_1,l_2}
r_{l_1} r_{l_2} \langle\phi_{l_2}\mid
\hat{Q}(k,k^\prime) \mid\phi_{l_1}\rangle
\overline{\langle\phi_{l_2}\mid
\hat{Q}(k,k^\prime) \mid\phi_{l_1}\rangle}.\nonumber\\
\label{Bound1}
\end{eqnarray}

The right-hand-side of the above inequality is identified with
the sum of the expectation values of $H_{\cal M}$ 
in the two states,
\begin{equation}
\Psi_1 = \sum_{l=1}^D r_l \psi_l^\uparrow
\otimes \bar{\psi}_l^\downarrow,\>\>\>
\Psi_2 = \sum_{l=1}^D r_l \phi_l^\uparrow
\otimes \bar{\phi}_l^\downarrow,
\end{equation}
where $\bar{\psi}_l$ and $\bar{\phi}_l$
are the complex conjugates of $\psi_l$ and
$\phi_l$, respectively. 
To see this, we note that
$\hat{T}_\sigma$ is hermitian
and $\{\hat{Q}_\sigma(k,\>k^\prime)\}$ are {\it real} in the 
representation chosen. A straightforward substitution establishes
that inequality (\ref{Bound1}) is equivalent to
\begin{equation}
E_0(2N + 1) \ge \frac{1}{2}
\langle\Psi_1\mid H_{\cal M} \mid\Psi_1\rangle +
\frac{1}{2} \langle\Psi_2\mid H_{\cal M} \mid\Psi_2\rangle.
\label{variational}
\end{equation}

>From Eqs. (\ref{number_op}), we see that, by construction,
$\Psi_1$ and $\Psi_2$ are wavefunctions in the subspaces
$V(N_\uparrow=N_\downarrow=N+1)$ and 
$V(N_\uparrow=N_\downarrow=N)$, respectively.
It is easy to verify that they are also normalized,
\begin{equation}
\langle\Psi_1\mid\Psi_1\rangle =
\langle\Psi_2\mid\Psi_2\rangle 
=\sum_{l=1}^D r_l^2 =1.
\end{equation}
Therefore, we may regard $\Psi_1$ and $\Psi_2$ as variational
wavefunctions for a system of $2N+2$ and $2N$ particles, respectively.
By the variational principle, we obtain,
\begin{equation}
E_0(2N + 1) \ge \frac{1}{2} E_0(2N + 2) +
\frac{1}{2} E_0(2N).
\label{Bound2}
\end{equation}
In other words, $\Delta_P\ge 0$. To finish our proof, we
need to show that inequality (\ref{Bound2}) is actually
{\it strict}.

Suppose Eq. (\ref{Bound1}) holds as an equality, 
we must then have, 
\begin{equation}
\langle\psi_{l_2}\mid \hat{C}_k^\dagger \hat{C}_{k^\prime}
\mid\psi_{l_1}\rangle
=\overline{\langle\phi_{l_2}\mid \hat{C}_k^\dagger \hat{C}_{k^\prime}
\mid\phi_{l_1}\rangle},
\label{equality}
\end{equation}
for all terms in (\ref{Energy}) with 
$r_{l_1}\ne 0$, $r_{l_2}\ne 0$ and $g(k,k')> 0$.
To show that this is not possible, let us consider
a subset of terms with $k=k'$ and $l_1=l_2$. 
Now, $\hat{C}_{k}^\dagger\hat{C}_{k^\prime}=
\hat{C}_{k}^\dagger\hat{C}_{k}=\hat n_k$. 
Applying Eqs. (\ref{Constraint1}) and (\ref{Constraint2}) to 
$\Psi_0(2N+1)$, we obtain
\begin{equation}
\sum_{k} \sum_{l=1}^D
r_l^2\langle\psi_l\mid \hat n_{k}
\mid\psi_l\rangle = N + 1
\label{psi_N+1},
\end{equation}
and 
\begin{equation}
\sum_{k} \sum_{l=1}^D
r_l^2\langle\phi_l\mid \hat n_{k}
\mid\phi_l\rangle=N.
\label{phi_N}
\end{equation}
(Note that the expectation value of $\hat n_k$ is real in
any given state.)
Since there are equal number of terms in the two sums, one must have
at least one pair which violate Eq. (\ref{equality}).
Consequently, when $g(k,k')>0$ for all pairs of admissible states
with $k=k'$, inequality (\ref{Bound2}) must be strict, i.e.,
$\Delta_P$ is strictly larger than zero.

Our proof is accomplished. {\bf QED}.

\subsection{Lower bound}
  
In the above proof, we only showed
that the parity parameter $\Delta_P$ is positive and nonzero.
Actually, when $g(k,\>k^\prime)=g>0$ is a constant for
pairs of the admissible states, we can further derive
a positive lower bound for $\Delta_P$. For this purpose,
we notice that inequality (\ref{Bound1}) and the ensuing inequality
(\ref{Bound2}) are derived
by replacing terms of the type $ab$ in the second sum
in Eq. (\ref{Energy}) with 
$\left(\mid a\mid^2+\mid b\mid^2\right)/2$. Taking into account the
fact that $E_0(2N+1)$ is real, we can express 
the error caused by this procedure in the form
$\mid a-b^\ast\mid^2/2$ for each term in the sum.
For simplicity, let us only estimate the error produced
by handling terms with $k=k^\prime$ and $l_1=l_2=l$. For such terms,
$\hat{C}_{k}^\dagger\hat{C}_{k}=\hat{n}_k$. Combining
Eqs.~(\ref{Energy}), (\ref{Bound1}) and (\ref{variational}), we obtain
the following inequality,
\begin{equation}
\Delta_P \ge \frac{g}{2} \sum_k \sum_{l=1}^D 
r_l^2 \big| \langle\psi_l\mid
\hat{n}_k \mid\psi_l\rangle -
\langle\phi_l\mid \hat{n}_k\mid\phi_l\rangle
\big|^2. 
\label{Bound3}
\end{equation}
Furthermore, by using the Cauchy-Schwarz inequality
$\left|\sum_n a_n b_n\right|^2\le\left(\sum_n\mid a_n\mid^2\right)
\cdot\left(\sum_n\mid b_n\mid^2\right)$,
we obtain
\begin{equation}
\Delta_P \ge \frac{g}{2} {{\left|\sum_k \sum_l
r_l^2 \left(\langle\psi_l\mid
\hat{n}_k \mid\psi_l\rangle -
\langle\phi_l\mid \hat{n}_k
\mid\phi_l\rangle \right)\right|^2} \over
{\sum_k\sum_l r_l^2}}.
\label{Bound4}
\end{equation}
>From Eqs. (\ref{normalization}), (\ref{psi_N+1}) and (\ref{phi_N}),
we obtain finally,
\begin{equation}
\Delta_P\geq \frac{g}{2N_E}.
\label{Bound5}
\end{equation}
Here again $N_E$ is the number of single-particle admissible states
in the relevant energy range.

\section{Discussion and Conclusions}

In this paper, by applying a generalized version
of Lieb's spin-reflection positivity technique, we 
established the existence of a nonvanishing parity parameter
$\Delta_P$ in an ultra-small superconducting grain and
derived a positive lower bound for $\Delta_P$ in a special case.
Our study complements previous analytical work in the weak coupling 
regime\cite{Matveev} 
and numerical investigations\cite{Mastellone,Berger}.
The bound (\ref{Bound5}) may seem to be too crude for practical
purposes. However, we emphasize that it is rigorous and its
validity does not require the coupling constant $g$ to be small.

An interesting open question is how pair-breaking scattering processes
influence $\Delta_P$ in the ultra-small grain limit.
This question has not been addressed explicitly in
previous theoretical studies 
\cite{Janko,Delft,Smith,Matveev,Braun1,Braun2,Mastellone,Berger,Balian}
based on the BCS Hamiltonian
(\ref{Hamiltonian1}), which does not contain any pair-breaking interaction.
In contrast, experimental work by Ralph, Black, and Tinkham
\cite{Ralph,Black} suggests
that pair-breaking due to spin-orbit interactions
may be quite strong in real systems. 
Since our method relies only on certain global symmetries of
the Hamiltonian, qualitative answers to the above 
question can be obtained in some cases.
To be more precise, let us consider separately
the following two situations.

(I) {\it Scattering processes that preserve spin symmetry}. 
In such a process, the spin of the scattered electron is not flipped. 
A familiar example is the scattering of electrons by a nonmagnetic impurity
described by the Hamiltonian,
\begin{equation}
H^\prime = \sum_{k_1\neq k_2} \sum_\sigma
\left(t(k_1,\>k_2) c_{k_1,\>\sigma}^\dagger c_{k_2,\>\sigma}
+ C.\>C.\right),
\label{Scattering Hamiltonian}
\end{equation}
where the scattering matrix
$\left(t(k_1,\>k_2)\right)$ is Hermitian. Apparently, $H^\prime$
is pair-breaking but it preserves the spin of the scattered electrons.
We notice that the combined Hamiltonian $H=H_{\cal M}+H^\prime$ can be formally
thought to be an ``extended {\it negative}-$U$ Hubbard model'' defined
on a lattice in the {\it single-particle eigenstate space}. (In the literature
of the strongly-correlated fermion models, such a lattice is called
a ``superlattice''\cite{Nagaoka,Tian2}). 
Such a model can be treated in a similar way
as in Ref.\cite{Tian1}, where 
the binding energy of the negative-$U$ Hubbard model
on a {\it real-space} lattice was discussed.
By following the proof of theorem 1
in Ref.\cite{Tian1} and the proof of the above inequality (\ref{Bound5}),
without further ado, one easily finds that the main result
of this paper is still valid. Thus, the scattering of electrons
by nonmagnetic impurities does not destroy
superconductivity in a small superconducting grain.

(II) {\it Scattering processes that do not preserve spin symmetry}. In such
a process, the Cooper pair is broken and the spin of the scattered electron
is also flipped by the interaction. Examples include
scattering by a magnetic impurity and spin-orbit interactions. 
While the former process definitely
suppresses superconductivity in the superconducting grain, the origin
and effect of the latter interaction have not been fully understood
\cite{Ralph,Black}. Due to the spin-flipping process, the total electron
spin of the system is no longer a good quantum number. Consequently,
our approach based on Lieb's spin-reflection positivity is not applicable
to these systems. To analyze the effects of such interactions on
the superconducting grain in a mathematically rigorous way,
more sophisticated techniques are needed. Further research on this
issue is currently under way.

\vspace{10pt}

  {\bf Acknowledgments:} One of us (G.S.T) would like to thank
the Croucher Foundation for financial support and the Physics
Department, Hong Kong Baptist University for their hospitality.
This work is also partially supported by the Chinese National
Science Foundation under grant No. 19574002.

\end{multicols}

\end{document}